\documentclass[aps,prl,floatfix,twocolumn, twoside, footinbib]{revtex4-1}
\usepackage{epsfig} 
\usepackage{amsmath}
\usepackage{amssymb}
\usepackage{mathrsfs}
\usepackage{amsthm}
\usepackage{enumerate}
\usepackage{siunitx}
\usepackage{float}
\usepackage{graphicx}
\usepackage{gensymb}

\usepackage{booktabs}

\usepackage{fancyhdr}
\usepackage{lastpage}
\begin{document}
\title{Fabrication of A Dual Gated Mirror Symmetric Twisted Trilayer Graphene Device to Study Superconductivity}
\author{Ahmed Shaikh, Phanibhusan Singha Mahapatra, Eva Y. Andrei}
\date{\today}
\affiliation{Department of Physics and Astronomy, Rutgers University, New Brunswick, NJ, USA }

\begin{abstract}

Though research on graphene by itself has waned, the interest in moir\'e materials, materials made with stacked layers of graphene with a rotational twist between the layers, has exploded in popularity. These layered devices show a key feature, flat bands. Flat bands localize electrons, which in turn leads to the expression of correlated states such as Mott insulators, superconductivity, and more. A key property of these devices is that their 2D nature allows us to tune them in situ, effectively allowing us to change the device's electronic properties. This powerful ability greatly reduces the time and money required to study superconductivity. The superconductivity in these systems seems to be similar to high-temperature superconductors such as cuprates, giving us a path towards studying high-temperature superconductivity. The fabrication of these devices is nontrivial, and thus we detail one general way to create these layered devices to give maximal tunability.

\par\vspace{1em}
\small\textbf{Note:} This document is an undergraduate honors thesis and has not undergone peer review.
\end{abstract}
\maketitle 

\section{Introduction}
Since the discovery of graphene in the early 2000s, the field of 2D materials has grown exponentially. This is especially true after it was discovered that moir\'e patterns, interference patterns generated by a misalignment of any two repeating patterns such as those of atomic lattices, fundamentally changes the band structure of the materials involved  \cite{Andrei2021}. This was first accidentally discovered in 2009 at Rutgers in Professor Eva Andrei's lab when looking at graphite with scanning tunneling microscopy (STM) \cite{Andrei2021}. The observation of Van Hove singularities in the band structure and their dependence on the twist angle gave an exciting new way to study correlated electron interactions. Furthermore, the realization that the geometry of 2D materials, with all their atoms accessible, lends itself to tuning of the electrical properties via a variety of techniques launched the field of moir\'e materials and 2D materials to greater heights. 

The discovery of the magic angle, a special twist angle initially discovered for twisted bilayer graphene (TBG), and its production of flat bands which host correlated electron states, further cemented the role of twisted graphene as a useful, interesting, and highly tunable platform for studying correlated states  \cite{cao2018unconventional}. The observation of superconductivity in twisted graphene was exciting as it seemed to mirror the superconductivity of unconventional high-temperature superconductors, such as cuprates. The key difference being that, unlike cuprates, these Van Der Waals heterostructures could be adjusted in situ, significantly reducing the cost and time required to research superconductivity. Since the discovery of superconductivity, other correlated states such as Chern insulators, strange metal phase, as well as exotic phenomena such as quantum anomalous Hall states, orbital magnetism, and more have been observed \cite{Andrei2021}. Many different moire stacks have been studied, including stacking many different layers at different angles, and even using other 2D materials such as MoS$_2$ to create specialized materials. Most exciting for us is the rich and wonderful system of mirror symmetric twisted trilayer graphene (MSTTG or TTG). This system hosts robust superconductivity and provides high tunability, which is key for probing these correlated states. In this paper, we show how to physically fabricate these devices as well as some preliminary measurements on one of these devices.

\section{Background}

\subsection{Motivation}
\begin{figure}
    \centering
    \includegraphics[width=\linewidth]{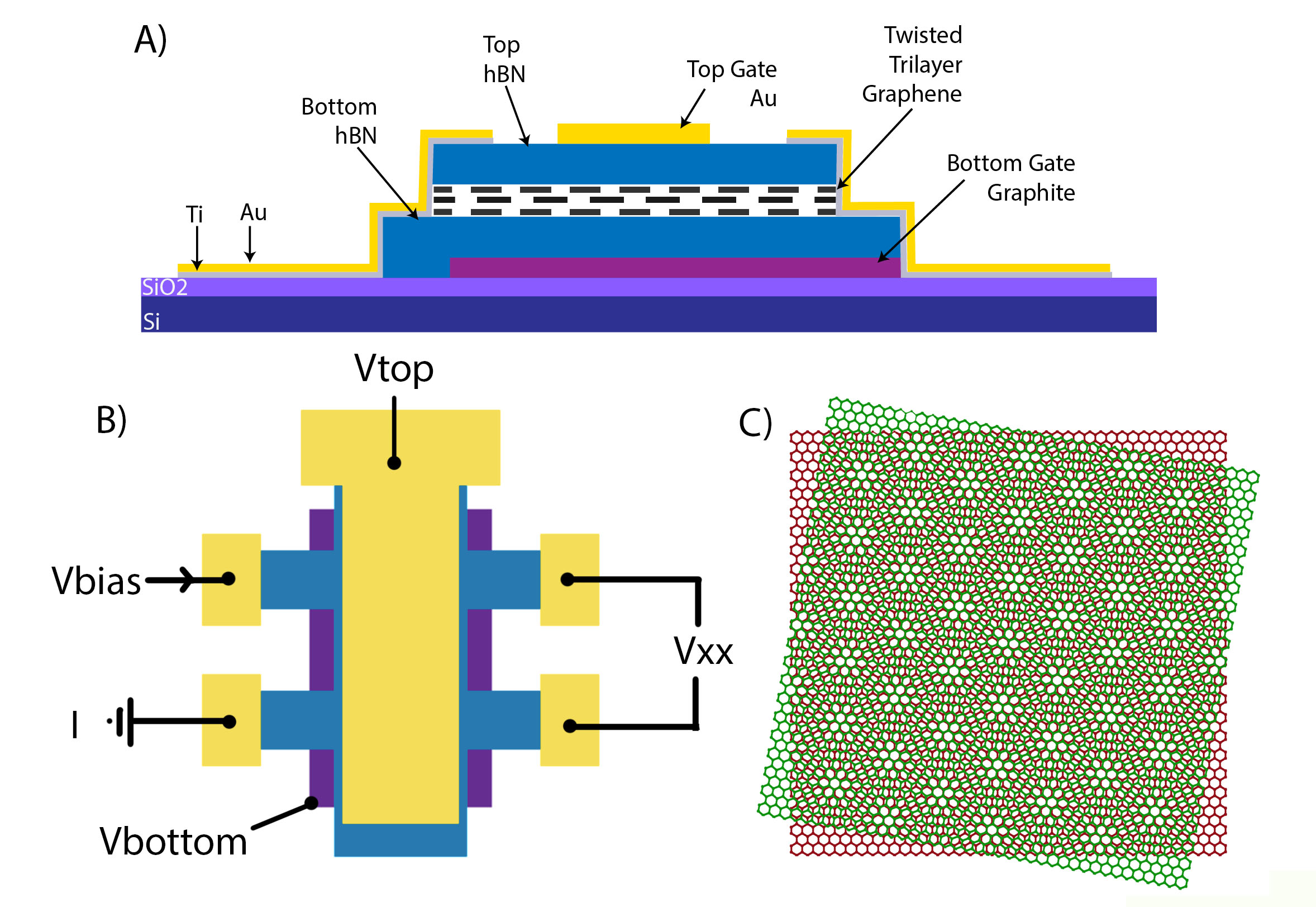}
    \caption{Schematic of the device structure. A) Side view of the device with materials labeled. B) Top view of the device with contacts labeled. C) Moire pattern emerging from the rotational misalignment of graphene layers.}
    \label{fig:schematic}
\end{figure}

Mirror symmetric twisted trilayer graphene is made up of three layers of monolayer graphene with the middle layer twisted at an angle $\theta$. The twist in the TTG creates a large-scale repeating moir\'e pattern which modifies the band structure, Figure \ref{fig:schematic}. The length scale of the moir\'e pattern is approximately proportional to $1/\theta$. Thus, for smaller angles, the moir\'e length scale is large and vice versa. For specific angles, called the magic angles, the band structure shows flat bands \cite{Khalaf2019}. Flat bands are a key feature because they localize electrons, leading to correlated states such as Chern insulators and superconductivity. For MSTTG, the first two magic angles are theoretically predicted to be $\theta = 0.414\degree$ and $\theta = 1.57\degree$ \cite{Khalaf2019}. The smaller magic angle is harder to achieve in actual devices, as small angles require more precise actuators, and also because graphene tends to relax to a more energetically favorable configuration, making small twist angles unstable. 

Previous work by other groups on MSTTG shows a rich phase diagram containing superconducting states flanked by correlated insulating states for specific filling values of the moir\'e cells \cite{Park2021}. MSTTG shows superconducting domes, which are highly dependent on the displacement field through the sample and the filling \cite{Park2021}. Thus, this system is highly tunable in situ, as using a dual-gated setup provides independent control over the carrier density and displacement field. Though this system has been studied before, the nature of superconductivity in twisted graphene is still a mystery. It is unknown yet whether it is conventional or unconventional superconductivity, and there are many open questions about how it exactly arises in this system. Thus, this system is a valuable one to study.

\subsection{Device Structure and Materials}

\subsubsection{Materials and Device Schematics}
The fabrication of these devices has evolved quite rapidly, though some techniques have endured. These moir\'e materials and Van Der Waals devices are made up of stacked layers of materials such as graphene, MoS$_2$, hexagonal boron nitride (hBN), and more. Physical exfoliation via micromechanical cleavage with the use of scotch tape, as was first done by Andre Geim and Konstantin Novoselov, remains one of the preferred ways of creating large, homogenous, and defect-free graphene flakes \cite{GeimExfol}. There are many methods for stacking these 2D layers, with wet chemistry techniques to dry transfer using viscoelastic materials. The technique we employ in this paper is the stack and tear method \cite{Kim2016}.

The device we will make is made of seven layers, using a variety of materials such as graphene, graphite, hBN, and gold. In the early days of these layered devices, graphene was studied on silicon substrates directly. However, silicon substrates are far from atomically flat, and the strain and disorder from the uneven surface changes the electronic properties. As such, the next evolution was studying suspended graphene sheets, laid over a gap etched into a silicon substrate \cite{Meyer2007}. However, such devices were impractical and sensitive, imposing strict limitations on device structure. The discovery of hBN as a substrate for graphene was groundbreaking. This is because hBN has the same hexagonal atomic structure as graphene, is completely insulating, and most importantly, is atomically flat. This allows for the creation of pristine, atomically flat stacks which preserve the electronic properties of graphene \cite{Dean2010}.

Our device is made up of MSTTG at a twist angle of 1.57\degree, sandwiched between two bulk layers of hBN. Then, on the bottom of this sandwich is a multilayer flake of graphite, which acts as the bottom gate, and on the top, we deposit some gold as the top gate. This entire device rests on a silicon substrate with a SiO$_2$ layer. We also deposit titanium-gold contacts to interface with the device. A side schematic is shown in Figure \ref{fig:schematic}. The top view (b) shows us the general structure of the device. We have contacts for the top and bottom gates, and we also have perpendicular bars placed across the main channel of TTG. These perpendicular bars are called hall bars, as we can use them to measure a hall voltage in the presence of a magnetic field. This setup, however, is more general than simply measuring the hall voltage. We can use any of the four contacts, labeled V$_{\text{bias}}$, I, and V$_{\text{xx}}$ to measure a wide variety of different voltage configurations. The standard measurement we do is called the four-probe resistance measurement. We apply a bias voltage to the bias contact, which results in a current passing through the device, which sets up a voltage difference between the two ends of the device, and then we use two other contacts to measure this voltage. From this, we can calculate the resistance.  Such a measurement scheme allows for the measurement of the device resistance without measuring the probe resistance. In our actual device, we attempt to make many hall bars to give us maximum flexibility in the different types of measurements we can take.

\subsubsection{Dual Gating}
Our device features two gates, hence the term dual-gated. As mentioned in the introduction, this dual gating allows for high tunability of the device. The device can be conceptually considered as a coupled capacitor setup with the shared plate of both capacitors being the TTG, which we connect to ground. The dielectric in the capacitors is the hBN, and the other two plates are the top and bottom gates. Thus, the end result is that we can control, independently, the carrier density, the amount of electrons or holes, in our TTG, and the displacement field, the electric field through the sample. The formula for the carrier density and the displacement field is shown in equations \ref{DisplacementField} and \ref{carrierDensity}. This allows us to explore the rich phase diagram of MSTTG by adjusting the top and bottom gate voltages, effectively allowing us to measure a "new" device at each carrier density. 

\begin{equation}
    D = \frac{(C_{top}V_{top Gate} - C_{bottom}V_{bottomGate})}{\epsilon_0}
    \label{DisplacementField}
\end{equation}

\begin{equation}
    n = \frac{(C_{top}V_{top Gate} + C_{bottom}V_{bottomGate})}{e}
    \label{carrierDensity}
\end{equation}

The $C_{top}$ and $C_{bottom}$ are the capacitances of the top and bottom hBN, and the $e$ is the charge of the electron. The $n$ is the carrier density and the $D$ is the displacement field through the sample.

\section{Device Fabrication}

Device fabrication can largely be broken down into three parts. The first is preparing the materials, which includes exfoliating hBN and graphene, preparing the stamp, and making a pre-patterned wafer. The preparation part is key, as work put into the preparation increases chances of success during the next part and ultimately determines how clean and large the device can be. The second part is stacking, in which the stack and tear process is used to create the layered device  \cite{Kim2016}. The final part is shaping the stack into a usable device via the use of lithography and some miscellaneous work on finishing up the device, such as wirebonding. We will go over these steps in detail now, with lots of figures to give a visual understanding of the process.

\subsection{Preparation of Materials}

The device fabrication process is highly dependent on the materials which are used in the process. A graphene flake that is the right size and shape, as well as a matching hBN flake with the correct dimensions and features, makes the stacking process more predictable and yields a cleaner device. As such, refining the exfoliation process and being highly selective in the choice of graphene and hBN flakes being used is key to success. The following section details the exfoliation process as well as the selection criteria for the flakes. The overall setup is also discussed in the last subsection.

\subsubsection{Substrates}

The entire fabrication process occurs on silicon substrates, which serve as the rigid backing that supports and allows us to handle the device on a macroscopic level. The substrate of choice is a silicon substrate with a 285 nm thick SiO$_2$ layer on the top surface. These substrates are the standard choice for 2D materials work, as substrates with an approximately 300 nm thick oxide layer provide high optical contrast for visually identifying graphene monolayers \cite{Wafer}. We cleave small square sections from a large silicon wafer via the use of a diamond scribe to score a line and the application of uneven pressure to allow for cleaving along the crystal structure. The resulting substrates are 1 cm by 1 cm squares of the Si/SiO$_2$ wafer. The substrate will henceforth be referred to as silicon substrate or substrate for the remainder of the paper.

The substrates are cleaned and prepared based on what they will be used for. All substrates undergo basic cleaning with acetone and isopropyl alcohol with the following procedure: Substrates are placed in an ultrasonic cleaner submerged in acetone for 1 minute, followed by drying off using inert nitrogen gas. Then these two steps are repeated with isopropyl alcohol. Finally, the substrates are placed on the hotplate at 350 \degree{C} for 15 minutes to burn off any remaining oils or organic residues. This level of cleaning is basic compared to other procedures, such as RCA cleaning, however, it is fast and provides a clean enough surface for hBN exfoliation and lithography of pre-patterned wafers.  

For graphene exfoliation, the substrates undergo plasma cleaning. After the basic cleaning, the substrates are placed in a vacuum chamber. A vacuum is pulled, after which oxygen is introduced. Our setup has pressure control but no flow rate control. We kept the oxygen pressure at roughly 55 mTorr pre-ionization and used 60 Watts of power to ionize the plasma. The substrates are exposed to the plasma for 35 seconds, which we found was the sweet spot for the level of surface attraction generated. The charged particles in the plasma ionize some of the oxide layer, which is beneficial for obtaining large graphene flakes. However, a surface attraction that is too high adversely affects the exfoliation process, leading to excess glue residue and tearing of the exfoliation tape.

\subsubsection{Exfoliation of Graphene}

\begin{figure}
    \centering
    \includegraphics[width=\linewidth]{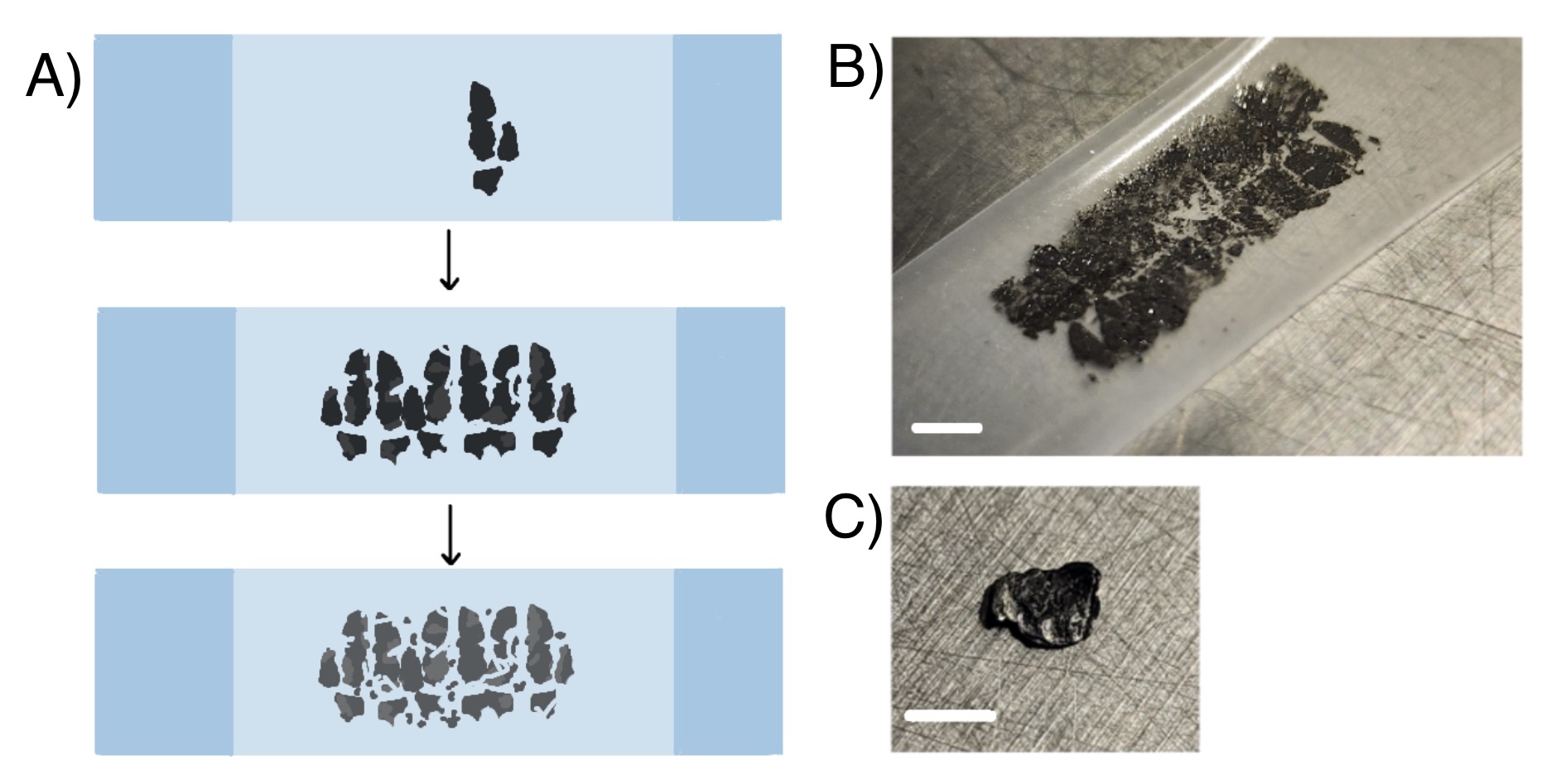}
    \caption{A) Schematic diagram of the preparation of the exfoliation tape. Preservation of the starting shape of the graphite crystal is attempted to maximize the size of the graphene. The graphite is spread out over the initial tape in the first two images. The third image depicts the tape after successive thinning operations. B) Photograph showing the final tape used for exfoliation. The density and thickness of the graphite are optimal for exfoliation. Scale bar is 1 cm. C) Example of a starting graphite crystal is shown. Scale bar is 1 cm.}
    \label{fig:exfo}
\end{figure}

\begin{figure*}
    \centering
    \includegraphics[width=\linewidth]{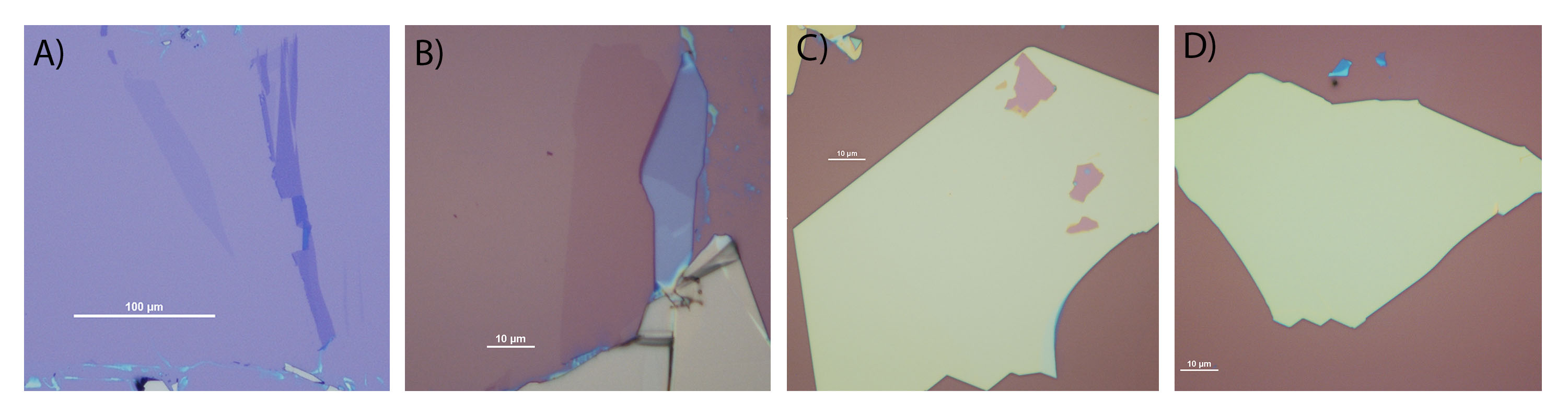}
    \caption{Optical microscope images of ideal and non-ideal graphene flakes and ideal top and bottom hBN. A) Ideal clean graphene flake with about 20 $\mu$m thickness and 100+ $\mu$m of length and no bulk graphite around it. Scale bar is 100 $\mu$m. B) Non-ideal graphene flake. Large and clean but connected to bulk. Scale bar is 10 $\mu$m. C) Ideal top hBN flake with a flat edge of length 70+ $\mu$m. Mostly clean and no cracks, and the right color/thickness. Scale bar is 10 $\mu$m. D) Ideal bottom hBN piece with clean surface and right thickness. Scale bar is 10 $\mu$m.}
    \label{fig:graphnhBN}
\end{figure*}

We use the scotch tape method for exfoliating graphene  \cite{GeimExfol}. We start with graphite crystals, which are typically 0.5 cm by 0.5 cm to 1 cm by 1 cm in size. We place the crystal on a piece of scotch tape (specifically Scotch Magic Greener tape), and fold the tape in half, allowing the sticky sides to come into contact. Then we unfold the tape, cleaving the crystal into two pieces which are now stuck on the tape. We repeat this folding and unfolding process systematically, spreading the crystal onto the tape. It is highly important to maintain the overall shape and integrity of the crystal during this process, as that will lead to larger graphene flakes. A schematic of this process is shown in Figure \ref{fig:exfo}.

Once the crystal has been effectively spread out over the tape, we now proceed to use an additional tape to thin down the graphite. We stick the two tapes sticky side to sticky side and press firmly to ensure good contact. Then, the tapes are peeled apart. This process is repeated with a fresh piece of tape, making sure to keep track of the order of the tapes, until the graphite has been thinned to the desired thickness. The key to achieving large graphene is to not thin the graphite down too much, an optimal tape is shown in Figure \ref{fig:exfo}. With each successive thinning, the crystals on the tape tend to break apart, leading to smaller flakes after exfoliating. We found that two to three rounds of thinning work best, though it depends on the thickness of the starting crystal. 

Once the tape has been prepared, we place plasma-cleaned substrates onto a hotplate at 100 \degree{C} and then place the tape onto the substrates on the hotplate. We gently press down on the back of the tape to maximize the contact between the graphite/tape and the substrate surface. We leave the tape on the substrates on the hotplate for 5 minutes, after which we remove the tape with the substrates still attached from the hotplate. We place the tape with the substrates onto a room temperature metal table and begin the peeling process. When peeling off the tape from the substrate, we make sure to peel the tape at a 180\degree angle from the surface of the substrate and peel at a slow but consistent speed. This yields the best results. We typically exfoliate on four substrates at a time using two tapes (2 substrates per tape). 

The final substrates are then observed under optical microscopy to look for graphene flakes. Monolayer graphene flakes tend to be only slightly darker than the surface of the substrate versus bilayers and multilayers. The ideal monolayer for a twisted trilayer device is a monolayer that is at least 75 $\mu$m long and around 20 $\mu$m wide. The monolayer should not be connected to bulk graphite, though if the monolayer is long enough, then it is acceptable. For the highest chance of success while stacking, look for a monolayer that is not surrounded by bulk graphite in at least one direction. Examples of ideal versus non-ideal monolayers are shown in Figure \ref{fig:graphnhBN}. It is important to be selective in choosing which flake to make a stack with, as that will greatly increase the chances of success in the stacking phase while also ensuring a cleaner device. For a typical four-substrate batch, we found that we usually ended up with two to three monolayer flakes, which met most if not all of these criteria.
 
\subsubsection{Exfoliation of hBN}

Exfoliation of hBN follows the same general steps as for graphene but with a few differences. For hBN exfoliation, we use semiconductor wafer processing blue tape. An hBN crystal is put onto a piece of blue tape and spread out by folding the tape onto itself. This time we use a radial folding pattern, folding along many different directions to spread out the hBN. This is because we are not looking for a monolayer but rather a bulk flake with a large straight edge, so the focus is on exfoliation yield/transfer onto substrate. Once the hBN has been spread out onto the tape, we thin the hBN using additional pieces of blue tape, in much the same way as for graphene. We can check if the tape is ready by inspecting the hBN on the tape under strong lighting and looking for the hBN to have a roughly purple color. Though a mix of colors is also usually fine.

We then take substrates that have been base cleaned and put the tape onto the substrates at room temperature. The tape is pressed down to ensure good contact. Then, after a minute or two, the tape is peeled back in much the same way as graphene. The key difference during the peeling process is that the speed should be much slower for the hBN tape. Then we examine the substrate under optical microscopy. The ideal hBN flake is as large as possible, with no glue residue, or cracks, and a long flat edge. It is important to check for any cracks or potential cracks using the LUT function on the microscope software used, as any potential cracks will almost definitely expand during the stacking process. The hBN's flat edge should be at least three times the width of the graphene that will be used for the stack, since that leads to the cleanest cut and the smallest likelihood of the graphene rolling up. Also, the thickness of the hBN is a crucial component. We can gauge the thickness of the hBN by looking at the color\cite{Anzai2019}. In general, a greenish-yellow color is what we aim for, as that signifies approximately 80 nm thickness. A darker mustard yellow means that the hBN is thicker, while a bluer color means that the hBN is thinner. Avoid any blue colored hBN as they are too thin. Any hBN flakes that are clean, large, and the right thickness, but do not have a flat edge, can still be used for the bottom hBN. Examples of ideal top and bottom hBN flakes are given in Figure \ref{fig:graphnhBN}.

\subsubsection{Stamp, Coating, and Setup}

For the stack and tear method, we use a hemispherical stamp for the stacking. The stamp is a drop of Polydimethylsiloxane (PDMS) on a glass slide. The stamp is cured at 100 \degree{C} for half an hour, followed by oxygen plasma cleaning at 60 Watts for one minute. The stamp is then spin-coated with a few drops of Polyvinyl Alcohol (PVA) at 2500 rpm for one minute. The PVA thickness is not highly critical. We want a coating that is not too thick but also covers the entirety of the stamp. It's important to check for bubbles in the PVA coating after spin coating using optical microscopy, as bubbles near the center of the hemispherical PDMS stamp can cause problems when stamping down onto the flakes. PVA is water soluble, so it can be dissolved in water; typically,y boiling deionized (DI) water for five to ten minutes works. The stamp can then be re-coated and reused. The PVA coating provides the attractive force, which is strong enough to overcome the attractive force between the graphene and the silicon substrate. It also allows us to make our stacks at 100 \degree{C} which yields cleaner stacks. 

The stacking setup consists of an optical microscope with a 20x and 100x lens and a dual-stage setup. The top stage has three linear axes of control while the bottom has three linear axes and one rotational axis of control. The axes have sub-micron control. The stages are controlled via a LabVIEW program and a game controller. The microscope is connected to a computer and provides a live view along with drawing/annotating features, which are crucial during stacking. The bottom stage is also heated and can be adjusted. It features a vacuum-pumped hole which is used to hold the substrates down. The stamp is affixed to the top stage, with the PDMS/PVA part facing the bottom stage. This entire setup is in an argon-filled glove box for cleanliness and isolation from vibrations. 

\subsection{Stack and Tear Method}

\begin{figure*}
    \centering
    \includegraphics[width=\linewidth]{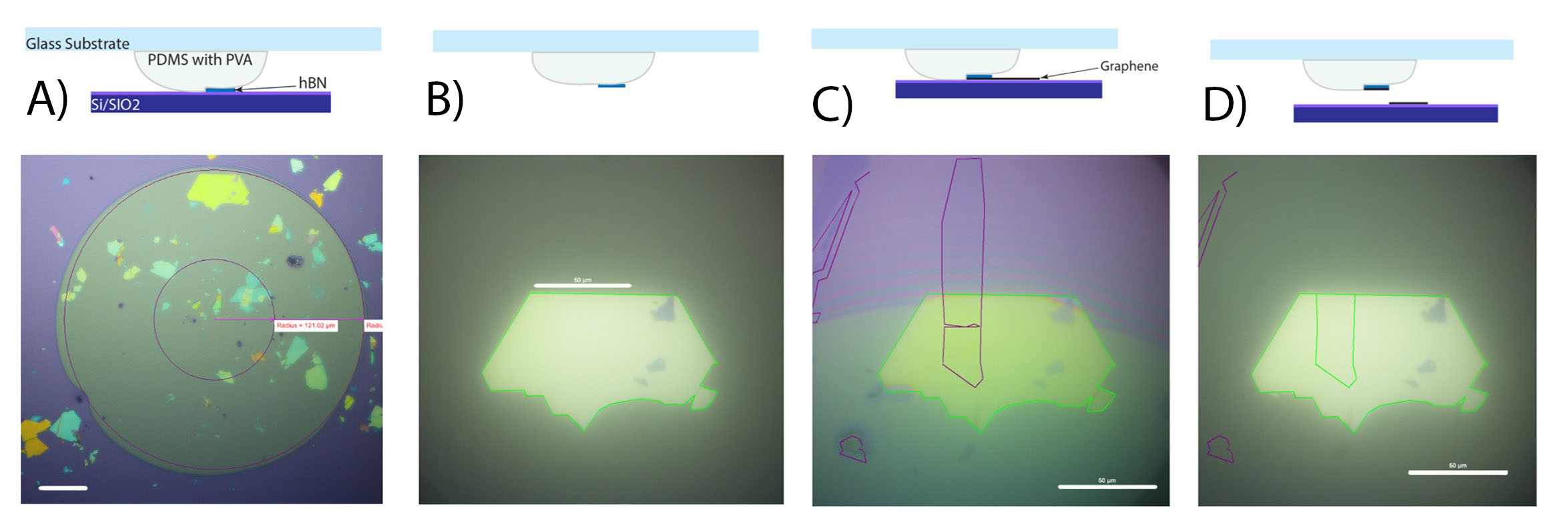}
    \caption{Schematic and optical microscopy images of one iteration of the stack and tear process. A) PDMS stamp with PVA touching down on substrate with hBN, touchdown point is roughly 300 $\mu$m away from the edge of the hBN. Scale bar is 100 $\mu$m. B) Image of hBN on the stamp after lifting up from the substrate. Scale bar is 50 $\mu$m. C) Stamp touching down onto graphene flake. The purple outline shows graphene, and the dark yellow is the hBN flake in contact with the graphene flake. Lighter yellow is the stamp in contact with the substrate surface. Scale bar is 50 $\mu$m. D) hBN with graphene on the bottom side, outlined in green, on the stamp after lifting from the substrate. Scale bar is 50 $\mu$m.}
    \label{fig:stackntear}
\end{figure*}

The stack and tear method relies on the difference between the adhesion of the materials at play to selectively pick up flakes  \cite{Kim2016}. For our twisted trilayer device, we need to pick up a top hBN, then three layers of graphene, followed by a bottom hBN, and a graphite gate. The pickup of the top hBN and the cutting of the graphene flake are the two main operations that require the most precision. The stacking process can be split into two parts: making the stack (Stacking), and transferring the stack onto a pre-patterned wafer (transfer). 

\subsubsection{Stacking}
First, we find a clean area of the substrate, which we will use to find the touchdown point of our stamp. We heat the bottom stage to 80 \degree{C}. Then we align the center of the hemispherical stamp with the substrate and touch down in the clean area of the substrate to locate the touchdown point. Once the touchdown point has been located, we lift up the stamp and move the bottom stage to position our hBN for pickup. It is important to place the cutting edge of the hBN facing away from the touchdown point, as shown in Figure \ref{fig:stackntear}. This is because we will use the fact that the expanding edge of the stamp will come from behind the cutting edge to press down onto the graphene up to the cutting edge of the top hBN. This ensures that only the hBN comes into contact with the graphene layer and not the stamp, allowing us to cut the graphene. It is likewise important to not pick up the hBN too close to the touchdown point of the stamp, as we have very little control over how large an area the stamp will spread over when it initially makes contact with the substrate. To get the most amount of control, we have found that placing the cutting edge of the hBN approximately 300 $\mu$m away from the touchdown point works best. Once everything is aligned, we can press down and past the edge of the hBN flake as shown in part A of Figure \ref{fig:stackntear}. Then we can lift up the stamp, and the hBN should get attached to the stamp, leaving the substrate behind. If this does not work, we usually increase the temperature by 5 \degree{C} and try again. The PVA is near its glass transition temperature, so small temperature changes greatly affect the behavior of the PVA.

Once the top hBN has been picked up, we can move the top stage out of the way and replace the bottom substrate with the substrate that has the graphene. We increase the temperature of the bottom stage to 100 \degree{C}. We then find the graphene flake we will use, and outline the graphene flake in our microscope software, at 100x magnification. This will allow us to align the graphene flake to the hBN, as seeing the graphene through the PDMS and PVA coating at 100x magnification is nearly impossible. It is also important to outline surrounding features to use as a reference for alignment. We then move the top stage to focus the microscope onto the hBN, and move the bottom stage down to keep the stamp away from the graphene. The hBN can then be aligned with the drawn outline, and we can now focus again on the graphene, moving the stamp up so that the hBN does not block our view. Now, we can slowly bring the top stage down until the stamp makes contact. Once the two stages are sufficiently close so that both the hBN and substrate are in focus, we switch to a speed of approximately 0.01 $\mu$m/s to give the most amount of control. We slowly press over the hBN, after having made sure of the alignment, and press the stamp down until the stamp has spread up to the edge of the hBN as shown in part C of Figure \ref{fig:stackntear}. 

Then we let the hBN remain in contact with the graphene flake for 5 minutes. After which we can lift up the top stage at a speed of approximately 0.1 $\mu$m/s. The graphene, which was in contact with the hBN, should be stuck to the hBN, and the graphene, which was on the substrate, should remain on the substrate, having been cut at the line where the edge made contact. We do this at 100x magnification to ensure proper alignment, which is key for picking up the next two graphene layers. This is in contrast to picking up the top hBN, which we do at 20x magnification to see the touchdown point and stamp spreading. We also use the outline of the graphene to keep track of where the layer we picked up on the hBN is, as there is no other way to tell.

We then repeat this process of outlining, aligning, pressing down, and lifting up to pick up the next two layers of graphene and then the bottom hBN and graphite gate. Some key things to note are that we use an ionizing fan pointed at the stamp roughly half a meter away in between graphene layers to reduce the chance of the graphene rolling up. Additionally, we rotate the bottom stage by +1.8\degree after picking up the first graphene layer and by -1.8\degree after picking up the second graphene layer. The magic angle for MSTTG is 1.55\degree, however, we use 1.8\degree as we have found that the graphene layers tend to relax by 0.25\degree. Also, the entirety of the stack, after the first hBN pickup, is done at 100 \degree{C} and 100x magnification. Picking up the bottom graphite gate can be tricky as the stack has many layers at that point, obscuring the view. It is essential to align the bottom gate with the twisted trilayer region to make a functioning device. Also, the choice of graphite is quite simple. We look for an evenly thick graphite flake, which has a dark purple color, indicating it is a many-layer flake rather than full bulk. We also look for a gate that can cover the entire twisted trilayer region. At the end of this process, we have a 6-layer stack on our PDMS/PVA stamp, made up of top hBN-TTG-bottom hBN-graphite gate. 

\subsubsection{Transfer}

\begin{figure}
    \centering
    \includegraphics[width=\linewidth]{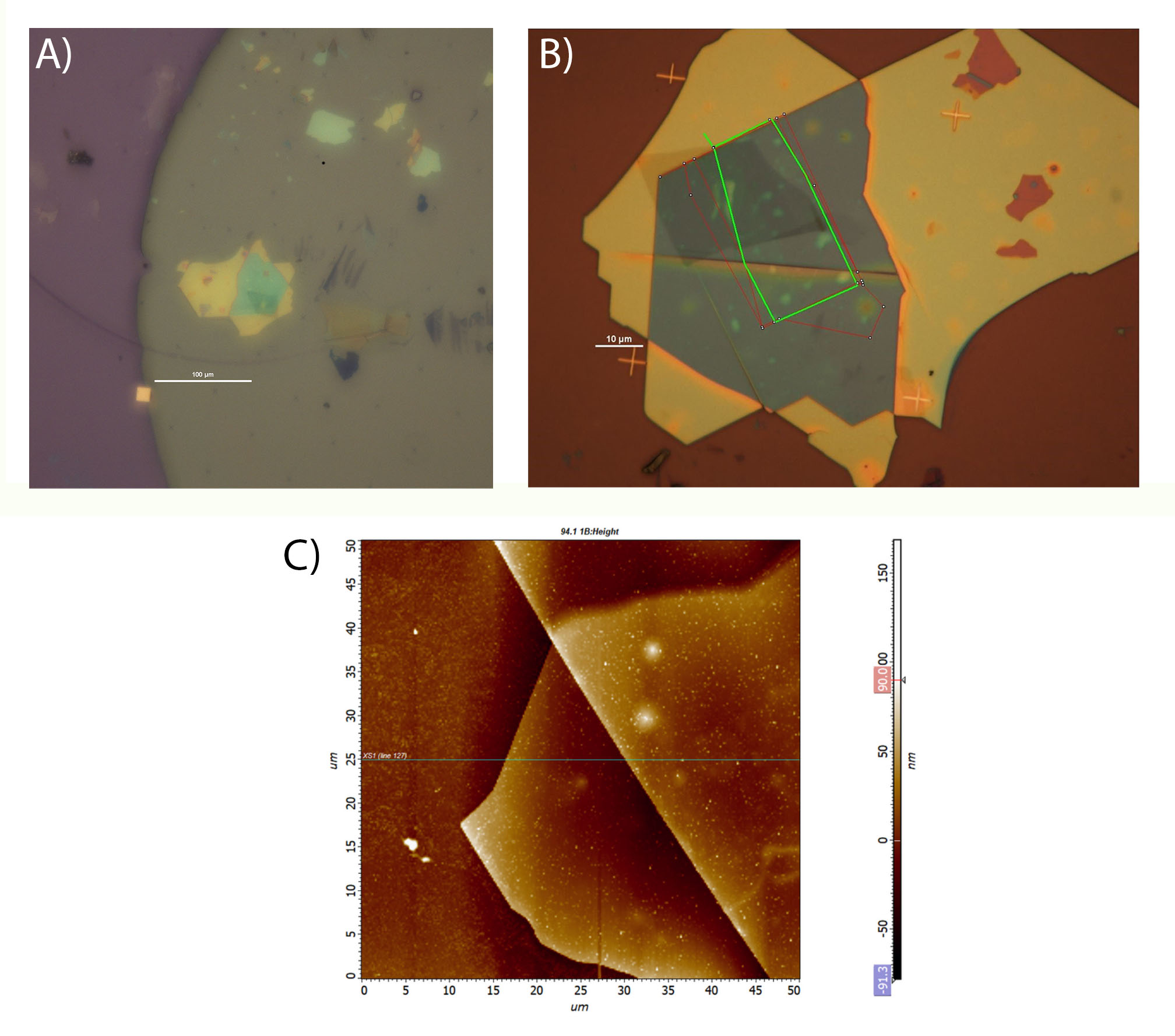}
    \caption{A) Stamp in contact with pre-patterned chip at 160 \degree{C} for transfer. Scale bar is 100 $\mu$m. B) Transferred stack on pre-patterned chip with twisted trilayer region (green outline) centered between four gold cross-shaped reference marks. Scale bar is 10 $\mu$m. C) Atomic Force Microscope image of the edge of the stack to determine the thickness of hBNs. The bottom hBN has a thickness of 77.5 nm $\pm$ 0.5 nm and the top hBN has a thickness of 85.6 nm $\pm$ 0.5nm. Bottom tick marks are 5 $\mu$m.}
    \label{fig:transfer}
\end{figure}

Once the stack has been completed, it needs to be transferred from the stamp onto a pre-patterned substrate. The pre-patterned substrate we use is a silicon substrate that has undergone basic cleaning and has had a grid of gold cross-shaped signs lithographed onto it, along with 3 large square gold pads on each side of the grid, which will serve as the contacts for the gates and hall bars later on. The transfer process is done outside the glove box as our setup inside the glove box is temperature-limited.

We start by removing the stamp from the glove box and mounting it to a manual top stage. We place the pre-patterned chip on a fixed heated platform and heat the substrate to 160 \degree{C}. We then use the microscope to align our stack with the area between four adjacent cross-shaped gold reference markers. We then slowly bring the stack into contact with the pre-patterned chip and leave it in contact for 6.5 minutes as shown in part A of Figure \ref{fig:transfer}. After which, we lift the top stage and ensure that the PVA has essentially melted and stuck to the pre-patterned chip. Then we transfer the stamp-stack-substrate sandwich to a small dish with DI water that has been heated to 160 \degree{C}, and we place the stamp-stack-substrate sandwich in the dish with the substrate side down and the stamp facing up. After 5 to 10 minutes, the PVA will dissolve and the stamp will release from the substrate. We then perform a basic cleaning of the substrate and look under the microscope to ensure that the stack has been properly transferred as shown in part B of Figure \ref{fig:transfer}. We image the finalized stack using an atomic force microscope (AFM) to check the thickness of the top and bottom hBN. The bottom hBN has a thickness of 77.5 nm $\pm$ 0.5 nm and the top hBN has a thickness of 85.6 nm $\pm$ 0.5nm. This stack has small dots, which are PVA residues that should not affect the stack processing much.

\begin{figure*}
    \centering
    \includegraphics[width=\linewidth]{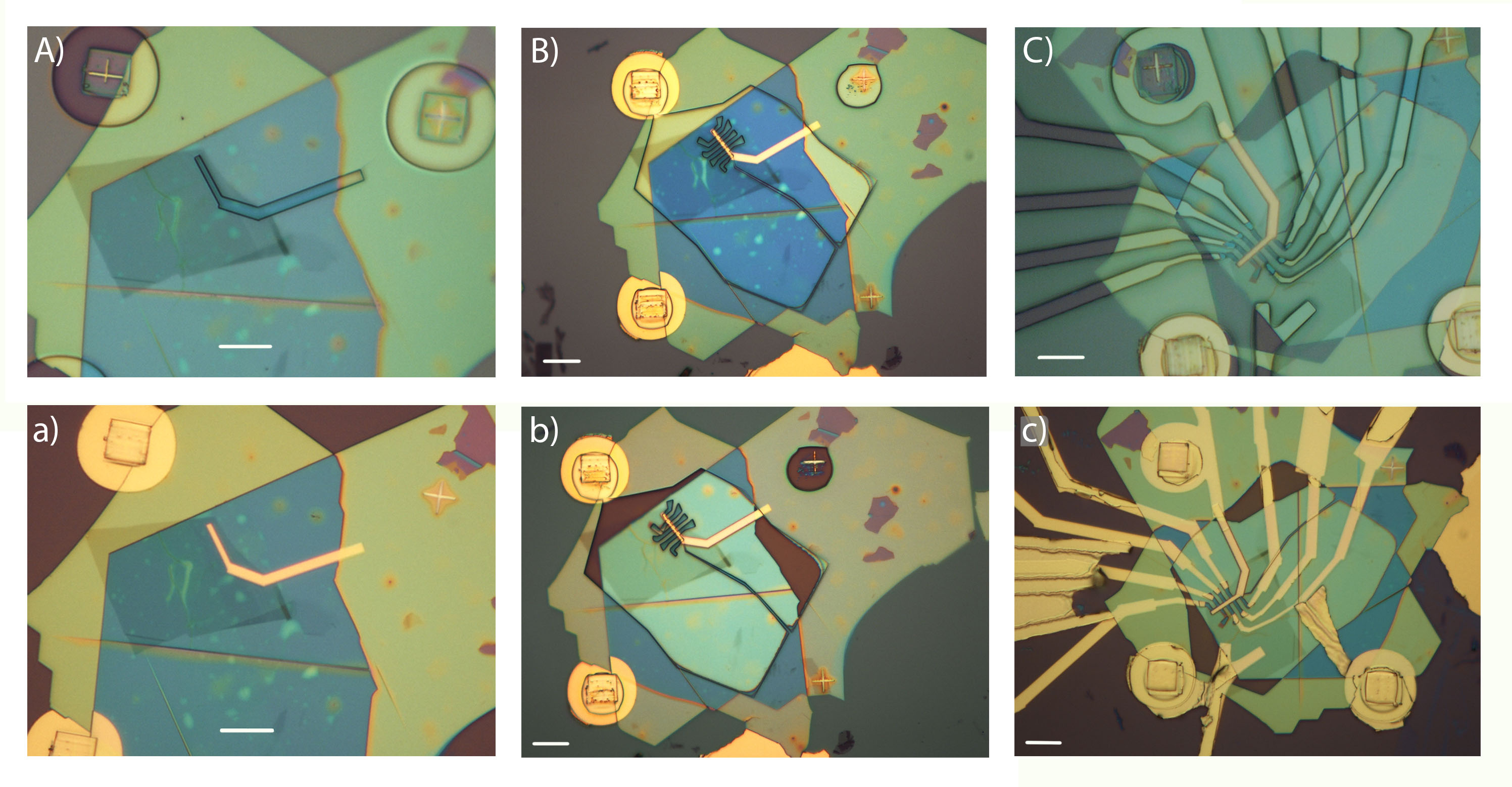}
    \caption{The lithography process shown step by step with an image of the PMMA mask and the corresponding feature developed via deposition of metals or reactive ion etching (RIE). A) PMMA mask of the top gate with the area chosen for its lack of bubbles and for its position within the twisted trilayer region. a) Gold deposited for the top gate. B) PMMA mask for defining the region that will be etched away via RIE. b) Stack after having been etched away and before the PMMA is dissolved. C) PMMA mask for the electrodes, which will be run to the top and bottom gates as well as the hall bars, connecting them with the square gold pads on the pre-patterned chip. c) The stack after the gold was deposited. This is after a second attempt using a different mask from the one in C. Three of the contacts have been shorted together due to problems with the gold deposition and lithography process. All scale bars are 10 $\mu$m.}
    \label{fig:litho}
\end{figure*}

\subsection{Lithography}

The final step in creating a usable device is to shape the stack with lithography. Lithography follows a simple step-by-step process, which is repeated to create various features, though the steps are always the same. We use electron beam lithography (EBL) to make our devices  \cite{Chen2015}. It starts with the layout of the traces and features on a computer using the gold reference marks to align an image of the stack to the pattern in the software. The gold reference marks are also used by the EBL machine to find its position and orientation. The chip with the device is spin-coated with Polymethyl Methacrylate (PMMA). The chip is then placed in the EBL machine, where the machine uses electrons to draw out the pattern. After which the chip is exposed to the developer, which removes the areas that have been exposed to the electron beam and finishes developing the PMMA. Then, the chip can undergo metal deposition or reactive ion etching (RIE) using CHF$_4$ in the same setup as the oxygen plasma device. After this, the PMMA is removed using a solvent, and the whole process is repeated to build up the device step by step. 

The entire lithography process is shown in Figure \ref{fig:litho}. For our device, we start by defining the channel by picking where the top gate will go. The area chosen was picked because it is bubble-free, is covered by the bottom gate, and is within the twisted trilayer region. We first deposit a small layer of titanium as it adheres to the hBN much better. We then deposit a thin layer of gold on top, on the order of a few nanometers. Then we mask off the hall bars as discussed in the device structure and materials sections, and etch away the top hBN and a bit into the bottom hBN. This etches away the twisted trilayer as well. The only regions where the twisted trilayer and top hBN remain after this step are the top gate and hall bar regions. This exposes the edge of the twisted trilayer along all the sides of the top gate and hall bars, while also exposing the edge of the bottom gate. We then lay titanium-gold contacts, connecting all the hall bars and both gates to individual pads along the edge of the chip. The metal contacts connect to the twisted trilayer and the bottom gate via the edge of these materials, thus, we call these edge contacts. Unfortunately, our setup for depositing gold ended up with off-gassing from the gold targets used to generate the gold mist, which deposits onto the device. As such, we ended up with contacts being shorted together after having to do part c of Figure \ref{fig:litho} twice. We also had an issue with the hall bar contacts showing a very high resistance, on the order of M$\ohm$. We suspect this was also due to the offgassing issue. These improper contacts were fixed with a gold patch repair at every hall bar connection after having fixed the metal deposition setup. 

Once lithography is done, we have only one final step to finish off the device and start using it to take data: wire bonding. We need a way to connect the gold pads to the dilution setup. We first mount the substrate with the device onto a removable sample stage with the use of GE varnish, and then we wire bond the gold pads to the contacts on the sample stage. The sample stage plugs into the dilution refrigerator, allowing us to control and interact with the device. With the wire bonding done, the device is complete and ready for experiments. 

\section{Experimental Setup}

The experimental setup consists of a few main components: a lock-in amplifier, two voltage sources, a dilution refrigerator, and the sample. The device is attached to the sample stage of the dilution refrigerator which is wired to a shunt box. The shunt box is connected to a lock-in amplifier, as shown in Figure \ref{fig:schematic} by the V$_{\text{bias}}$, a current of 1 nA is sent through the V$_{\text{bias}}$ line. The lock-in amplifier is also connected to the V$_{xx}$ to measure the voltage drop across the channel. This creates a four-probe resistance measurement, allowing us to remove the probe resistance. The top and bottom gates are connected to two voltage sources. Thus, we can keep one of the gate voltages constant and sweep through voltage values for the other gate while plotting the resistance. This allows us to do a preliminary check to see if the device superconducts or not. If superconductivity is achieved, then we can sweep through the voltages more systematically, maintaining constant carrier density while sweeping the displacement field or vice versa. The dilution fridge is initially cooled down using liquid nitrogen and then cooled down further by using liquid helium, allowing us to reach the sub-kelvin temperature scale (40 mK). 

\section{Data and Analysis}

Unfortunately, due to time constraints and problems with lithography, we were only able to complete one device out of the three stacks we constructed. As such, we were unable to take extensive data, though we were able to take some basic preliminary data on the one device that we managed to construct completely. We started by keeping the top gate grounded and sweeping through the bottom gate and vice versa to check the response of the device to the different gates at room temperature. 

\begin{figure}[h]
    \centering
    \includegraphics[width=\linewidth]{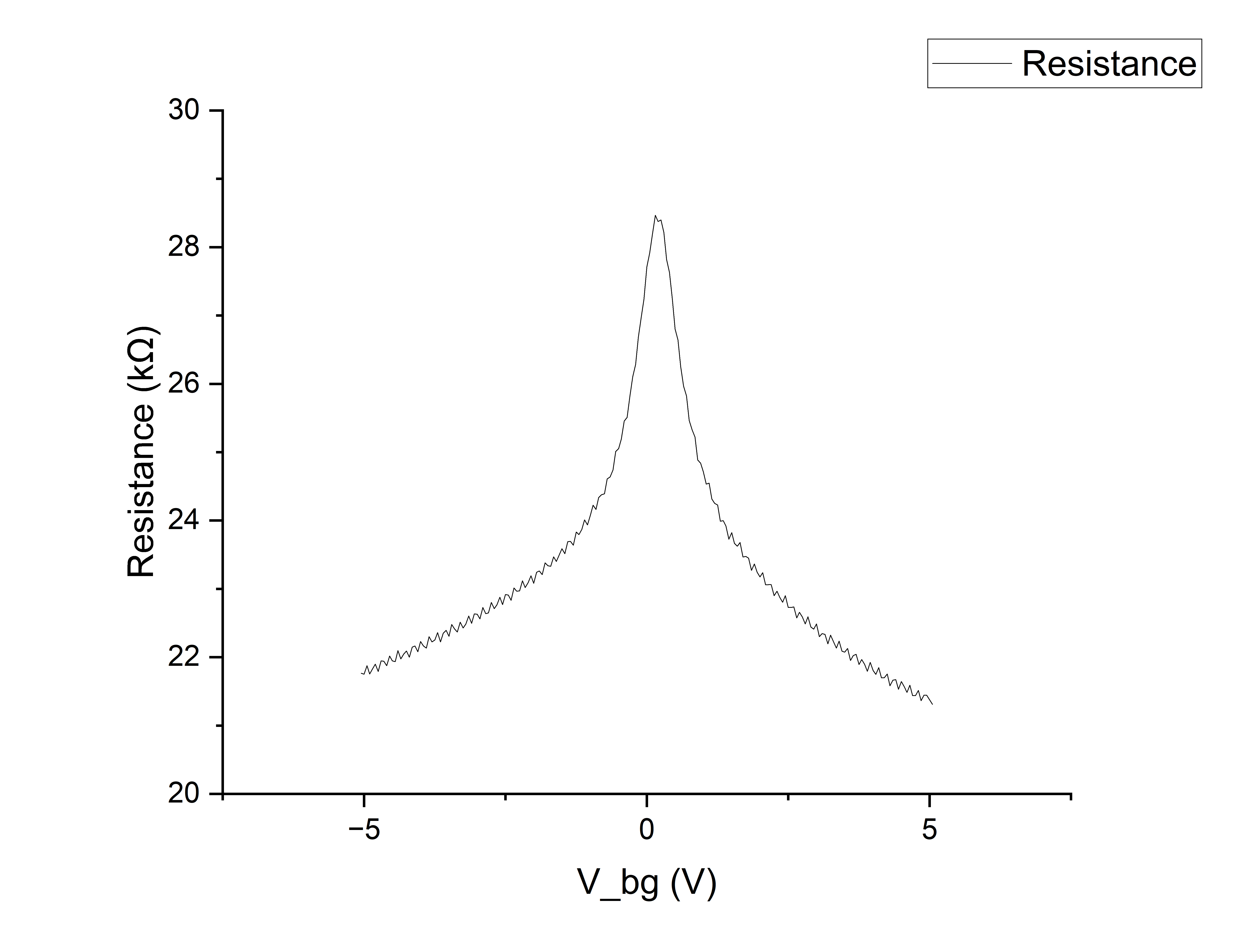}
    \caption{Plotting the resistance as a function of bottom gate voltage at room temperature, measured at 300K.}
    \label{fig:RT_bg}
\end{figure}

\begin{figure}
    \centering
    \includegraphics[width=\linewidth]{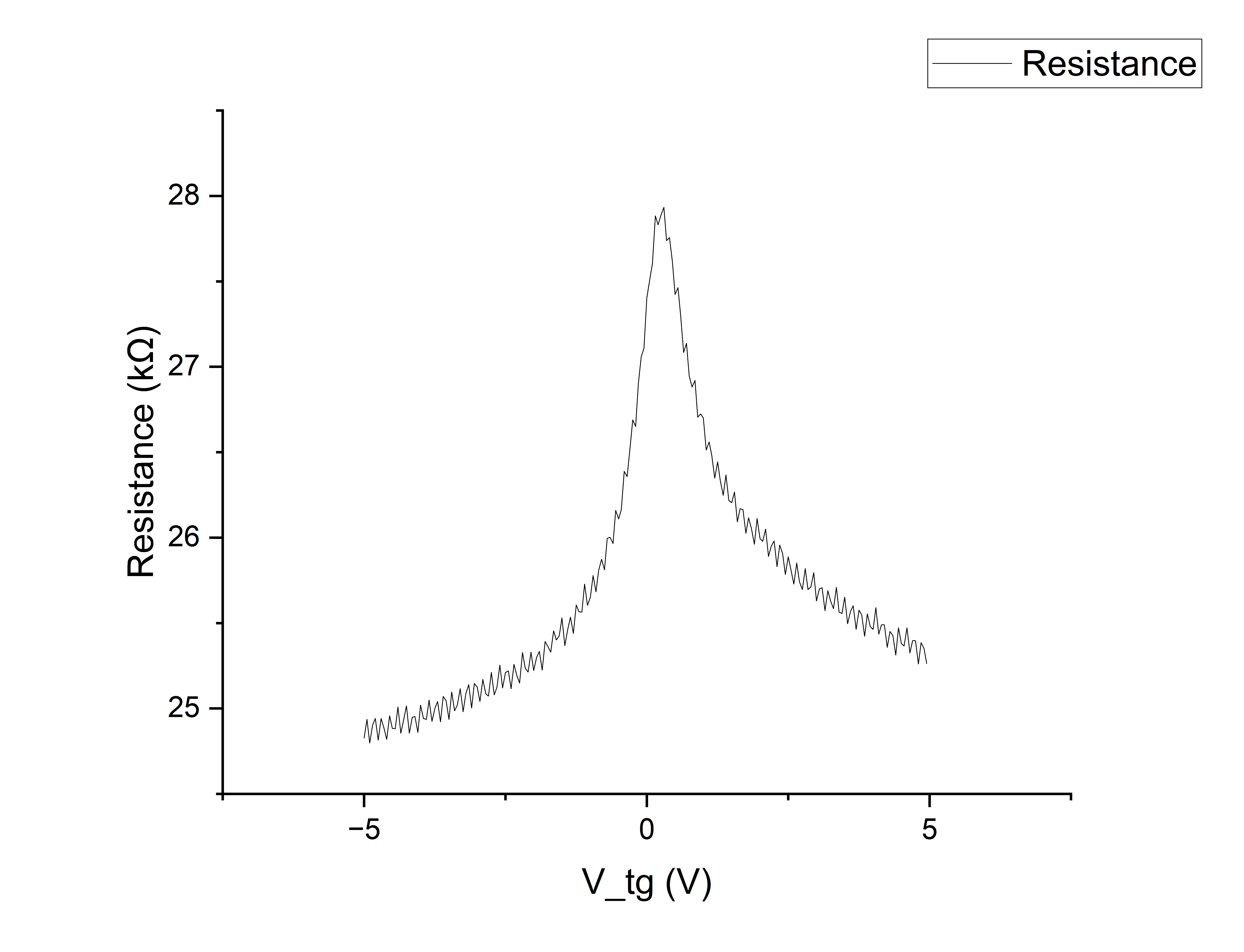}
    \caption{Plotting the resistance as a function of top gate voltage at room temperature,  measured at 300K.}
    \label{fig:RT_tg}
\end{figure}

From the resistance vs gate voltage sweeps, we can identify the Dirac point, the point where the two Dirac cones meet in the band structure. The resistance spikes because the density of states is low at this point. We see that the bottom gate, Figure \ref{fig:RT_bg}, seems to be more sensitive than the top gate, Figure \ref{fig:RT_tg}, at room temperature. Applying -5 volts to the bottom gate allows us to reach a resistance of below 22 k$\ohm$ where whereas applying -5 volts to the top gate only gets us to slightly below 25 k$\ohm$. Thus, we would need a much higher gate voltage on the top gate to reach the same response as the bottom gate. Since the response of the device to the gate voltages is proportional to the capacitance, which is proportional to the top and bottom hBN thicknesses, this leads us to conclude that the top hBN is thicker than the bottom, a conclusion that matches our AFM measurements. 

\begin{figure}
    \centering
    \includegraphics[width=\linewidth]{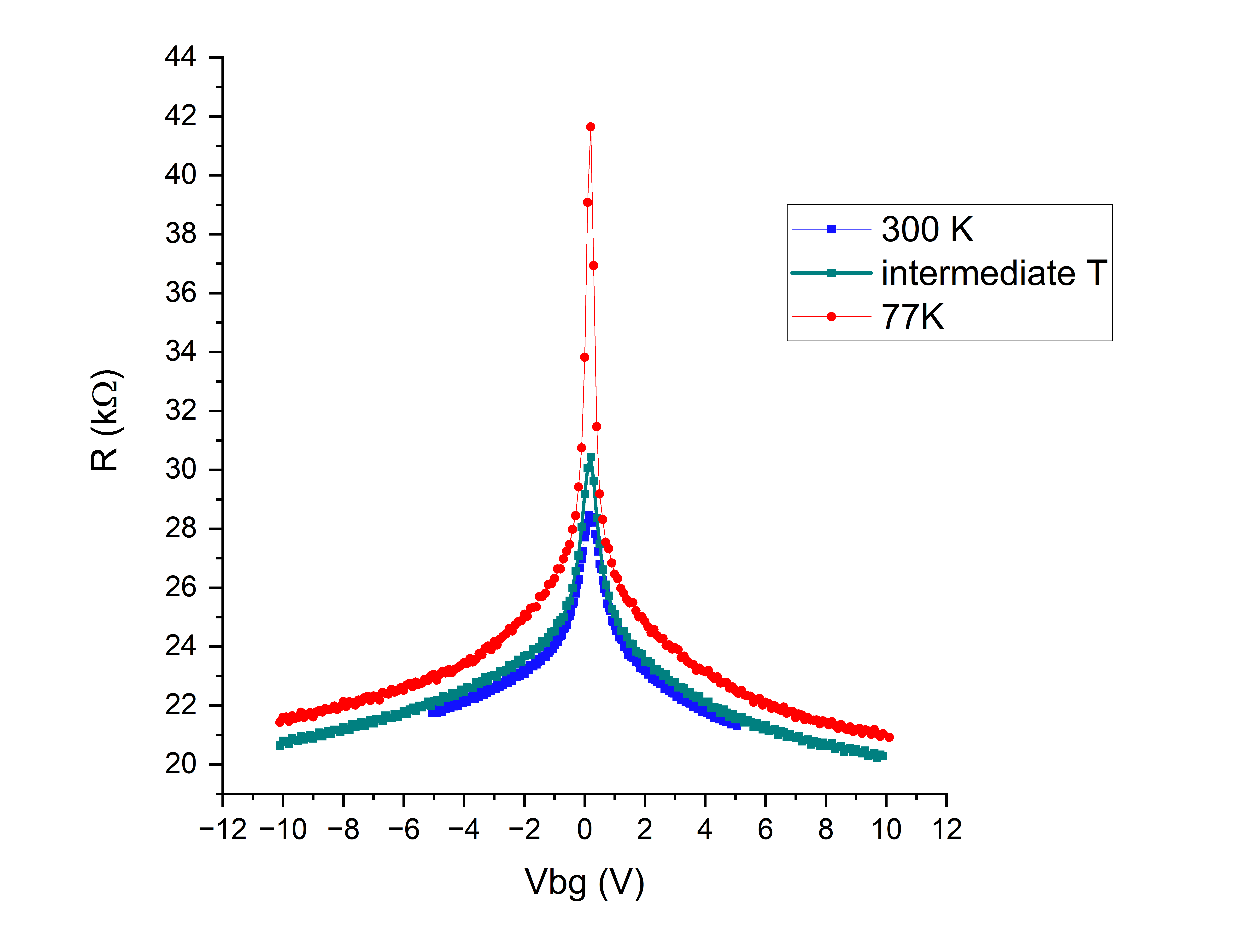}
    \caption{Plotting resistance as a function of bottom gate voltage for different temperatures between 300K and 77K.}
    \label{fig:temp}
\end{figure}

From here, we cooled down the device to 77K using liquid nitrogen, taking sweeps of the bottom gate voltage at room temperature, an intermediate temperature, and at 77K to check the device response as a function of temperature in a coarse-grained manner. This is plotted in Figure \ref{fig:temp}. As we see, the resistance of the device increases, and the Dirac point has a much higher peak at 77K than at room temperature. This matches what we expect for twisted trilayer, which hosts correlated insulating states that produce this inverse relationship between temperature and resistance. 

\begin{figure}
    \centering
    \includegraphics[width=\linewidth]{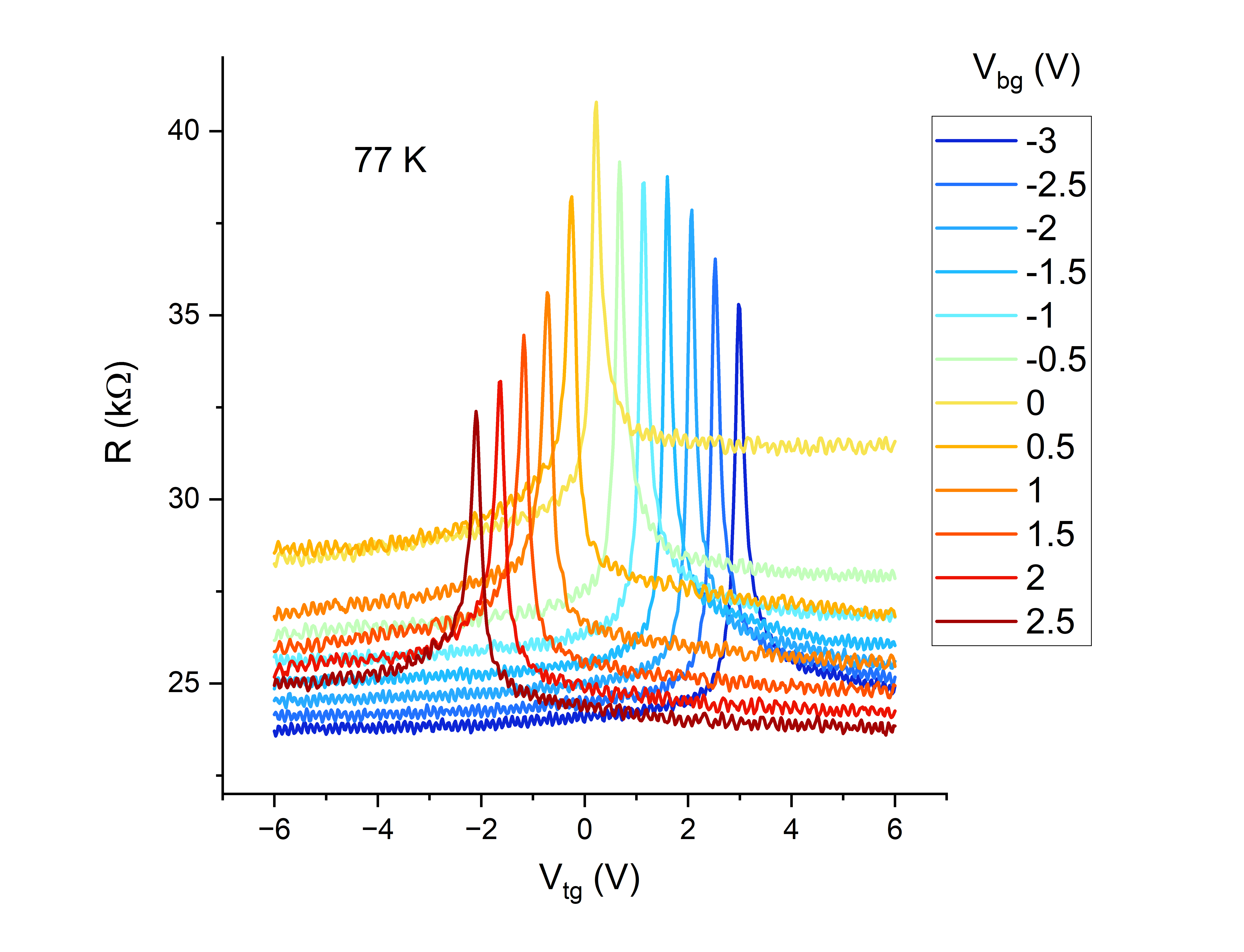}
    \caption{Plotting resistance as a function of top gate voltage for different bottom gate voltages from -3V to 2.5V, measured at 77K.}
    \label{fig:diracMovement}
\end{figure}

\begin{figure}
    \centering
    \includegraphics[width=\linewidth]{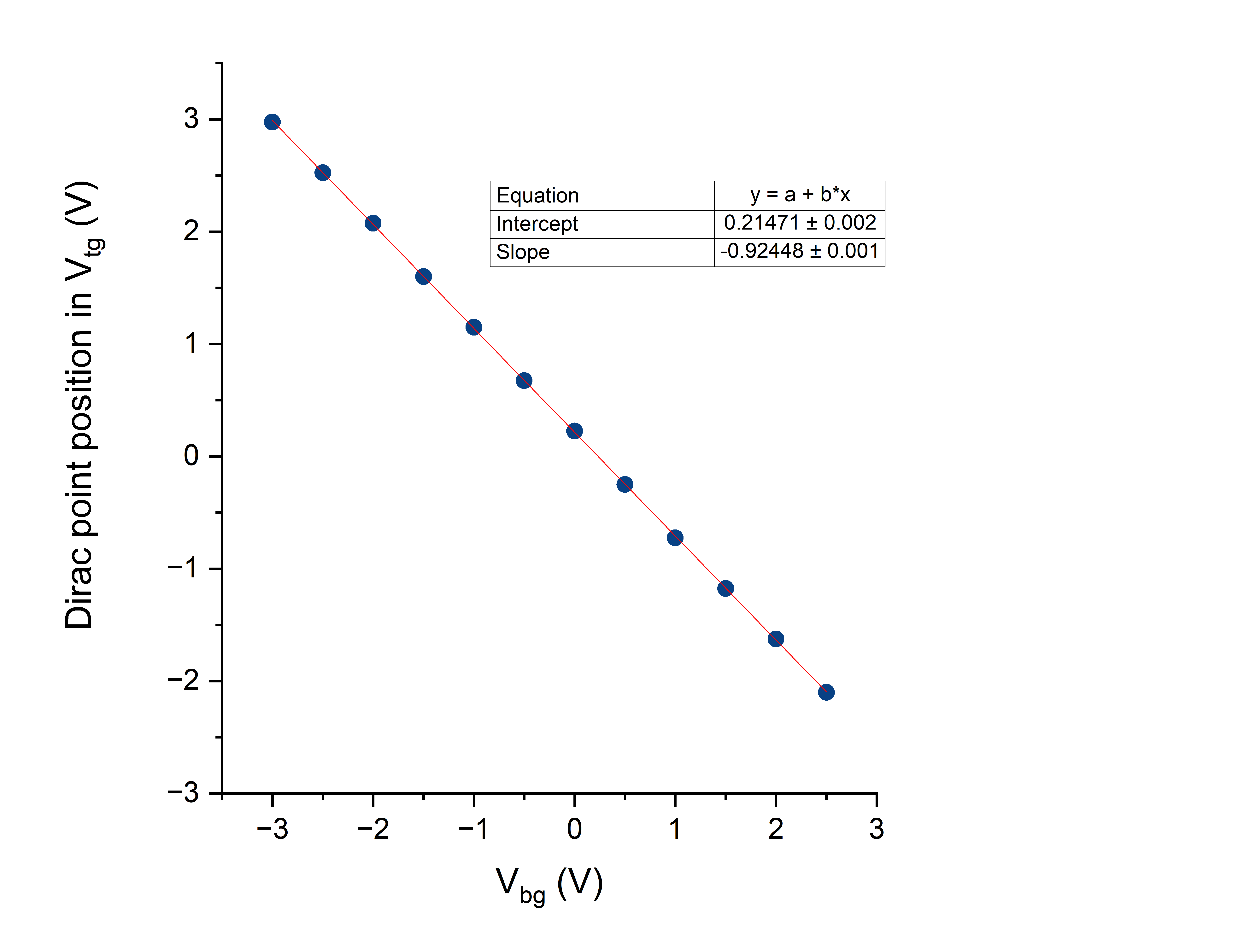}
    \caption{Plotting the position of the Dirac point in units of top gate voltage as a function of bottom gate voltage, measured at 77K. A linear fit has been applied to extract the slope of this relationship. The slope is approximately -0.92.}
    \label{fig:linFit}
\end{figure}

We can also check the ratio of the capacitances of the top and bottom hBN by checking the position of the Dirac point. For hBN of equal thicknesses, we expect that applying +V to the bottom gate should require applying -V to the bottom gate to reach the Dirac point. Thus, at 77K, we sweep through different top gate voltages and measure the resistance while keeping the bottom gate voltage the same during each sweep. This data is plotted in Figure \ref{fig:diracMovement}. We see something surprising here: the top gate requires a smaller voltage than the voltage applied to the top gate to reach the Dirac point. To see this more clearly, we can plot the position of the Dirac point in units of top gate voltage as a function of the bottom gate voltage. This relationship should be linear, and the slope of the line should tell us the ratio between the top and bottom hBN thicknesses. This is shown in Figure \ref{fig:linFit}. The data has been linearly fit, and the slope is -0.92. This tells us that the top hBN is thinner than the bottom, which contradicts our AFM measurements and room temperature data. One possible explanation for this is that our top gate is not fully gating our ttg. This is plausible since we had problems with the gold deposition on the hall bars. Perhaps we have a similar issue at the top gate, which is leading to non-ideal gating, thus lowering the response to the top gate voltage. 

Regardless, we continued with our testing and cooled down the device to liquid helium temperatures, around 300 mK. We then did the same sweeps as we did at 77K and 300K, and unfortunately, were not able to see superconductivity in this device. We think that one possible cause of this is that the twist angle is off from the magic angle of 1.55\degree. This is because the shape of the resistance vs bottom gate voltage graph, Figure \ref{fig:temp}, has a more peaked shape compared to data from a superconducting device that we had access to previously. Another reason could be the non-ideal gating, which is not allowing us to get to the right filling of the moire cell to reach superconductivity. Overall, the conclusion seems to be that this is a device-related issue and should be fixed if we do measurements on a new device.

\section{Future Work}

\begin{figure}[ht]
    \centering
    \includegraphics[width=\linewidth]{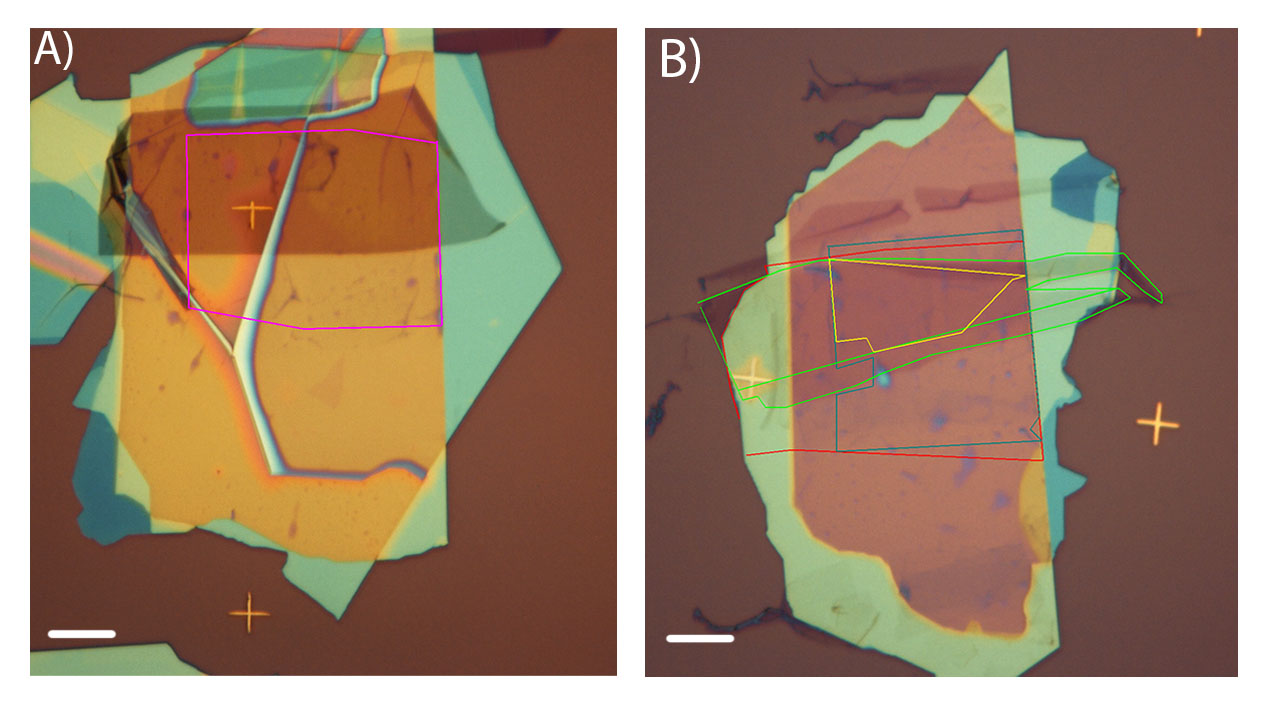}
    \caption{Optical microscope image of two additional stacks that were made alongside the first one. A) The purple outlined region is the twisted trilayer region. This stack has a crack in the bottom hBN, however, there is a lot of usable area, which we can make a device from. B) The yellow outlined region is the twisted trilayer region, with a clean region in the center, which can be used to make a device. Both scale bars are 10 $\mu$m.}
    \label{fig:stacks}
\end{figure}

Though we did not see superconductivity in this device, we have two other stacks that still need to undergo lithography. These stacks are much cleaner with fewer bubbles and impurities, as well as having better bottom gate alignment. These stacks are shown in Figure \ref{fig:stacks}. With the gold deposition setup fixed, we plan on doing lithography on these stacks to turn them into usable devices. Then we can cool them down in the dilution fridge and check for superconductivity. Also, having learned the device fabrication process in detail, we can now construct stacks relatively quickly. So if these stacks do not show superconductivity, which is unlikely, we can make a few more, adjusting for any systematic errors that we find. 

\section{Conclusion}

Though our device did not show superconductivity, we were still able to show how the dual gating works, and we were able to demonstrate the entire device fabrication process. We were also able to see a correlated insulating phase in our stack, which manifested as the inverse relationship between temperature and resistance. There is still a lot of work to do in finishing the lithography on the other two stacks. Once we have prepared those stacks, we can use our setup to see if they superconduct. After having tested one device, we have ironed out and tested our experimental setup and can be sure that it is adequate for probing these correlated states in our device. Though the fabrication process covered in this paper was to create a dual-gated twisted trilayer stack, the process itself is more general and can be applied to make any arbitrary 2D stacked device, even if the physical stacking setup is different, because the principles are still the same. 

\section{Acknowledgments}
I would like to thank my advisor, Prof. Eva Andrei, for her guidance and help with the project, and for providing me with the opportunity to work on this project. As well as Dr. Phanibhusan, a post-doc in our research group, who massively contributed to this project by teaching me the device fabrication process as well as helping me with the dilution fridge, taking data, and more. I would also like to thank Alexander Takhistov for showing me the stacking process and working with me to help me develop my skills in exfoliation of graphene and hBN. 

\section{ }
\bibliographystyle{apsrev4-1}
\bibliography{sources}
\end{document}